\documentclass[preprint]{aastex}

\shorttitle{Precise radial velocities of Polaris}
\shortauthors{Byeong-Cheol Lee et al.}

\begin{document}

\title{Precise radial velocities of Polaris: Detection of Amplitude Growth}

\author{Byeong-Cheol Lee\altaffilmark{1,2}, David. E. Mkrtichian\altaffilmark{3,4},
    Inwoo Han\altaffilmark{1}, Myeong-Gu Park\altaffilmark{2}, and Kang-Min Kim\altaffilmark{1}}

\altaffiltext{1}{Korea Astronomy and Space Science Institute, 61-1 Whaam-Dong
    Yuseong-Gu,  Daejeon 305-348, Korea; bclee@kasi.re.kr, iwhan@kasi.re.kr, kmkim@kasi.re.kr}
\altaffiltext{2}{Department of Astronomy and Atmospheric Sciences, Kyungpook National University,
    Daegu 702-701, Korea; mgp@knu.ac.kr}
\altaffiltext{3}{Astrophysical Research Center for the Structure and
    Evolution of the Cosmos, Sejong University, Seoul 143-747, Korea; davidm@sejong.ac.kr}
\altaffiltext{4}{Astronomical Observatory, Odessa National University,
    Shevchenko Park, Odessa,65014, Ukraine}

\begin{abstract}
We present a first results from a long-term program of a radial
velocity study of Cepheid Polaris (F7 Ib) aimed to find amplitude
and period of pulsations and nature of secondary periodicities.
264 new precise radial velocity measurements were obtained during
2004--2007 with the fiber-fed echelle spectrograph Bohyunsan
Observatory Echelle Spectrograph (BOES) of 1.8m telescope at
Bohyunsan Optical Astronomy Observatory (BOAO) in Korea. We find a
pulsational radial velocity amplitude and period of Polaris for
three seasons of 2005.183, 2006.360, and 2007.349 as 2K = 2.210
$\pm$ 0.048 km s$^{-1}$, 2K = 2.080 $\pm$ 0.042 km s$^{-1}$, and
2K = 2.406 $\pm$ 0.018 km s$^{-1}$ respectively, indicating that
the pulsational amplitudes of Polaris that had decayed during the
last century is now increasing rapidly. The pulsational period was
found to be increasing too. This is the first detection of a
historical turnaround of pulsational amplitude change in Cepheids.
We clearly find the presence of additional radial velocity
variations on a time scale of about 119 days and an amplitude of
about $\pm$ 138 m s$^{-1}$, that is quasi-periodic rather than
strictly periodic. We do not confirm the presence in our data the
variation on a time scale 34--45 days found in earlier radial
velocity data obtained in 80's and 90's. We assume that both the
119 day quasi-periodic,
noncoherent variations found in our data as well as 34--45 day variations found before
can be caused by the 119 day rotation periods of Polaris and by surface
inhomogeneities such as single or multiple spot configuration varying with the time.

\end{abstract}


\section{Introduction}

Polaris ($\alpha$ UMi, HIP 11767, HD 8890, HR 724 ) is one of the most famous
Cepheid variable stars.
In addition to its special location on the celestial sphere, Polaris has many interesting
astrophysical features.
It is a member of a triple system.
It is the brightest and the closest Cepheid variable with very low pulsational amplitude.
%It has been extensively observed since its
%variability was first discovered by Seidel in the middle of the 19th century,
%and extensive photometric and radial velocity measurement(RV) data have been accumulated.
It has been extensively studied over 1.5 centuries for the pulsational amplitude and
period changes using photometric and spectroscopic observations.
Perhaps the most remarkable feature of Polaris as a Cepheid variable is that the
period and amplitude of pulsation is changing very rapidly.
The pulsation period is rapidly increasing with the rate of about 4.5 s y$^{-1}$
(Turner et al. 2005 and references herein).
More interesting is the change of amplitude.
It is discovered that the pulsational amplitude has been decreasing dramatically
during the 20th century (Arellano Ferro 1983; Dinshaw et al. 1989).
So it was predicted that the pulsation of Polaris would completely stop by the end of
the 20th century.
However, Kamper \& Fernie (1998) notes that decline of radial velocity (RV) amplitude
has stopped abruptly.
Figure 5 of Hatzes \& Cochran (2000) shows the increase of amplitude,
but they did not state that explicitly.
Some recent photometric observations also indicate the same trend in the amplitude
(Davis et al. 2002; Engle et al. 2004).

\section{Observations and data reduction}

The new RV observations of Polaris were carried out during November 2004 to June 2007
using the fiber-fed high resolution (R=90,000) echelle spectrograph BOES (Kim et al. 2007)
attached to the 1.8m telescope at BOAO.
Using a 2k x 4k CCD, the wavelength coverage of BOES is 3600--10,500\,{\AA} with $\sim$ 80 spectral
orders in one exposure.
Observations were acquired through iodine absorption cell (I$_{2})$ to provide the precise
RV measurements.
Total of N=264 spectra were recorded;
the exposure time varied from 60 to 300 s depending on the sky condition
to get a typical S/N ratio of 250.
The extraction of normalized 1--D spectra was carried out using IRAF (Tody 1986) software package.
After extracting normalized 1--D spectra, the RV measurements were undertaken using a code called
RVI2CELL (Han et al. 2007) which was developed at BOAO.
RVI2CELL adopted basically the same algorithm and procedures described by Butler et al. (1996).
However we model the instrument profile using the matrix formula descried by Endl et al. (2000).
We solved the matrix equation using singular value decomposition instead of maximum entropy method
adopted by Endl et al. (2000).
With these configuration, we achieved the typical internal RV accuracy between 10 and 15 m s$^{-1}$
depending on the quality of spectra.

\section{Results}

Figure 1 plots the relative RV measurements for 2004--2007 season of observations.
The solid line is a zero-point adjusted trend for binary orbit calculated
according to the period of 29.59 years and orbital elements given by Wielen et al. (2000).
To study pulsations, we first removed the RV variation due to orbital motion.
Next, we applied  the Discrete Fourier Transform (DFT) analysis for
unequally spaced data to all de-trended 2004--2007 RV data.
We used for analysis the computer code PERIOD04 (Lenz \& Breger 2005).
The top panel in Figure 2 shows the resulting DFT periodogram  of all data.
We easily found a main frequency of $f{_1}$ = 0.251757 $\pm$
0.000008 c d$^{-1}$ ($P{_1}$ = 3.97208 $\pm$ 0.00013 days).
The whole 2004--2007 data phase diagram of Polaris phased with this period is plotted in Figure 3.
There are visible scatters, and two order values (near JD 2453902) exceed the accuracy of
our individual RV measurements ($\sim$ 10 m s$^{-1}$).
This scatter seems to be due to additional intrinsic RV variations to the dominant pulsation mode,
already known for Polaris (Dinshaw et al. 1989; Kamper 1996; Kamper \& Fernie 1998; Hatzes
\& Cochran 2000).
We removed the best fit sine-wave signal from dominant period from all the data string.
The DFT analysis of RV residuals is shown in second from top panel in Figure 2.
The highest peak at $f_{2}^{'}$ = 0.00799 $\pm$ 0.00010 c d$^{-1}$
($P{_2}^{'}$ = 125.1 $\pm$ 0.1 days) has an amplitude of 2K = 0.38 $\pm$ 0.02 km s$^{-1}$.

One might suspect that the secondary signal at $f'_{2}$ found in our 2004--2007
data residuals is due to removing dominant pulsations with fixed amplitude and period
which are actually varying during this time interval.
Here our main concern is to estimate the amplitude and period of the dominant pulsations
for short subsets as accurately as possible and remove it from RV variations in order to
study possible residual signals.
In determining the period and amplitude variation, the data were divided into three subsets:
Set 1, Set 2, and Set 3.
These subsets are marked in Figure 1.
The RV data  characteristics are given in second, third, and forth columns of Table 1.
Then, for each data set, we tried to find best fit frequency and amplitude for a dominant pulsation.
Figure 4 shows a dominant period fit to every subset of the data.
The r.m.s. residuals after the sinusoidal fitting of Set 1, Set 2, and Set 3 are 111 m s$^{-1}$,
149 m s$^{-1}$, and 63 m s$^{-1}$ respectively, much larger than the typical error of
RV data, 10--15 m s$^{-1}$.
It is, indeed, because there exists unmodelled RV variation in addition to the sinusoidal signal.
Table 1 shows the result of the period analysis: periods and amplitudes of pulsations found
for all three sets.
We applied the same period analysis to the most recent RV data published for Polaris.
The period and amplitude found during our re-analysis are also given in Table 1.

To study additional variability, the residuals after removing the best
fit dominant period and amplitude from each set were combined,
resulting in data string shown in Figure 5. As seen by eye, there is
still very strong $\sim \pm$ 350 m s$^{-1}$ and about 120 days time
scale variability. The shape of RV variability has the steep rise of
RV from negative to positive values and rapid drop back to negative
values. Such type of variability resemble closely the RVs variations
due to a contrast surface spot passing across the  visible disk in
some types (e.g. Ap) of stars. To get the accurate period the DFT
analysis was applied to the residual data. The DFT amplitude spectrum
is shown in the third panel from the top panel in Figure 2. The
largest 138 $\pm$ 8 m s$^{-1}$ peak is at $f_{2}$ = 0.00840 $\pm$
0.00003 c d$^{-1}$ (P$_{2}$ = 119.1 days). The residual data phased to
this period shown in Figure 6. The DFT of residuals after removing of
the secondary periodicity is shown in the bottom panel in Figure 2; it
does not show any significant peaks.

The check whether secondary periodicity may be due to
unrecognized aliasing problem and unfavorable  time sampling
of RV data (note, that the strongest sidelobes  of the spectral
spectral window function are 1 c d$^{-1}$ equally spaced)  we modeled the
dominant mode pulsations by mono-periodic sinusoidal signal having the
same time sampling as the original Polaris RV data. Note, that DFT of
this mono-periodic signal is actually the spectral window function
centered at $f_{1}$. We added to this mono-periodic signal the normally
distributed noise with amplitude of 138 m s$^{-1}$. The DFT analysis of these
artificial data is shown in  right panels (top and bottom) in Figure
6. As can be seen from the bottom panel, after removal of the
artificial signal, the  amplitude spectrum do not show any significant
signal at 0.008 c d$^{-1}$, confirming  that secondary 119-day periodicity we
found is not an artifact.

Figure 7 shows a century-long variations of the pulsational amplitude
and period of Polaris. As seen, our new data obtained for three
subsets of our 2004--2007 observations reveals that after a decade of
standstill the amplitude of pulsation is now rapidly increasing. We
can claim now safely that the era of amplitude decrease of Polaris was
finished in the end of 1980's and replaced with a new, quickly
uprising amplitude trend in the beginning of the 1990's.

\section{Discussions}

Historic change of the sign of amplitude variations in Polaris, that
we securely detected, doesn't yet have an analogy among other
Cepheids. Polaris is crossing the instability strip for the first time
and lies well in its center for fundamental mode pulsators or well
inside and near the hot bounds of instability strip for a first
overtone pulsators (Turner et al. 2005). We can suggest that switching
of amplitude change from decay to growth found in our observation is
not a direct result of evolution to a red border and the decrease of
the efficiency of the excitation mechanism, but might be relevant to
the unrecognized effect of mode interaction. We confirm the increase
of pulsational period too.

The detected long-term  characteristic period of 119 days is much
longer compared to the 9.75 days period by Kamper et al. (1984),
45.3 days by Dinshaw et al. (1989), 34.3 days by Kamper \& Fernie (1998) or
40.2 days by Hatzes \& Cochran (2000) reported in the 80's and 90's
data. We do not confirm the existence of any of the aforementioned
periods in our data. Dinshaw et al. (1989) argue that approximately
45-day period, that is not coherent, arises from one or more surface
features on Polaris carried across the disk by rotation.
Hatzes \& Cochran (2000) have found very reliable bisector variations
with the same period as in RVs and re-discussed three mechanisms of
the low amplitude residual RV variations in Polaris: a low-mass
companion, long-periodic nonradial stellar pulsations, or rotational
modulation by surface features.
They modeled the residual RV and
bisector variations for all hypotheses  - namely, existence of a
surface microturbulence, cool or hot spots and 45 day rotation period,
or the l=4, m=4 nonradial mode pulsations and found reliable RV
amplitude fit for all of them. However, the bisector span velocity for
all hypotheses had a unmodelled phase shifts, indicating that all of
these models do not satisfy completely observations.
Hatzes \& Cochran (2000) concluded, that among considered hypotheses
the nonradial pulsations has better agreement with modelling.

In our work based on long-term high-precision RVs, we do not confirm
any of the secondary period found earlier, but find clear 119.1 day
characteristic time of variations. This is a longest period of
intrinsic variations of Polaris found so far.

Summarizing all previous
investigation, we can claim that the secondary variations seen in
Polaris are not coherent on a long time-scales so we can firmly reject
first two hypotheses involving the companion and coherent
non-radial pulsations. Note, that period 119 days is too long compared
to a period of fundamental mode pulsations of Polaris and cannot
belong to normal acoustic modes.

We suggest two possible explanations of secondary radial velocity
variations:

a) The diversity in secondary periods found in Polaris is
likely the result from the rotational modulation of RVs by a single or
multiple surface spots and assume that the rotation period is about
119 days. In the case of multiple spots, the observed
periods about 40 days or about 34 days  are close to fractions
of 119 day period.  The rotational velocity of Polaris is
$vsini$ = 8.4 km s$^{-1}$ (Hatzes \& Cochran, 2000). The radius of Polaris
33 $\pm$ 2 $R_{\odot}$ was found by Turner et al. (2005) from the distance to
Polaris 94 $\pm$ 4 pc and the angular diameter 3.28 $\pm$ 0.02 mas given
by Nordgren et al.(2000). With these parameters and assumed rotation
period of 119 days, the equatorial rotation velocity is
$v_{\rm eq}$ = 14 km s$^{-1}$, that yields the  inclination angle of rotation axis
$i$ = $38^\circ$. The  projected rotational and expected equatorial
velocities of Polaris are in a quit good agreement with the
determined mean projected rotational velocities $\overline{v sin
i}$ = 15.8 $\pm$ 1.4 km s$^{-1}$ of a  sample of  eleven  F6-F7 Ib  supergiants
selected from the Ib supergiant list of De Medeiros et al. (2002).

b) We cannot exclude also that long-term cyclic variations of radial
velocities of Polaris is a results of oscillatory
variations of the mean radius of Polaris stochastically driven with
yet unrecognized physical mechanism.
Not excluded, that such a long-term, non-coherent and small radius
variations are intrinsic to many F type supergiant stars and Cepheids, but
in classical Cepheids are hidden in a large
amplitude of of the dominant mode pulsations and were not detected due
to limited RV accuracy and sparse time sampling of RV curves in
previous RV investigations. If true, Polaris might be a first
Cepheid star with well documented observational detection of such of
modulation. The another good example might be a F5 Ib supergiant
$\alpha$\, Per that show long-term (see Fig. 1 in Hatzes and Cochran
1995) variations. The detailed comparative precise RV study of a
sample of F-type supergiant stars including Cepheids can provide
additional constrains on the nature of long-term radius variations
detected in Polaris.

%{\bf For a F7Ib super-giant star like Polaris, and we cannot exclude
%the surface spots for Polaris.
%}

%We assume, that the period of 119.1 days found by us is actually the
%rotation period, and shorter periodicities found in previous studies
%were due to the multiple spots on the surfaces.
%The surface spots attributed to active regions are quite
%frequent phenomena on the convective surface of giant stars
%{\bf (e.g., Gray \& Brown 2006 and references herein)} but do not
%studied well in F-supergiant stars.

 The check of spot hypothesis and detailed analysis of
line profile and surface temperature variations in Polaris based BOES
observations is out-of the scope of the current work which is devoted
to the study of pulsational amplitude of RVs and will be presented in
our next paper.

\section{Conclusions}

First results of our long-term monitoring of Polaris show remarkable changes
in the amplitude and the period of pulsations that occurred at the end of 20th and
the beginning of 21st century.
The half-century-long pulsational amplitude decay was replaced with the rapid amplitude growth
while the growth of pulsational period continues.
We also detected the 119-day secondary RV variation that is about
three times longer that secondary periodicities reported before.
We concluded that 119 day variations are  not coherent on long
time scales and discussed possible nature of these variations.

\acknowledgements
BCL, DEM and MGP acknowledge the support from the Astrophysical Research Center for the Structure and
Evolution of the Cosmos (ARCSEC, Sejong University) of the Korea Science and
Engineering Foundation (KOSEF) through the Science Research Center (SRC) program.

\clearpage

\begin{figure}
\centering
\includegraphics[width=8cm]{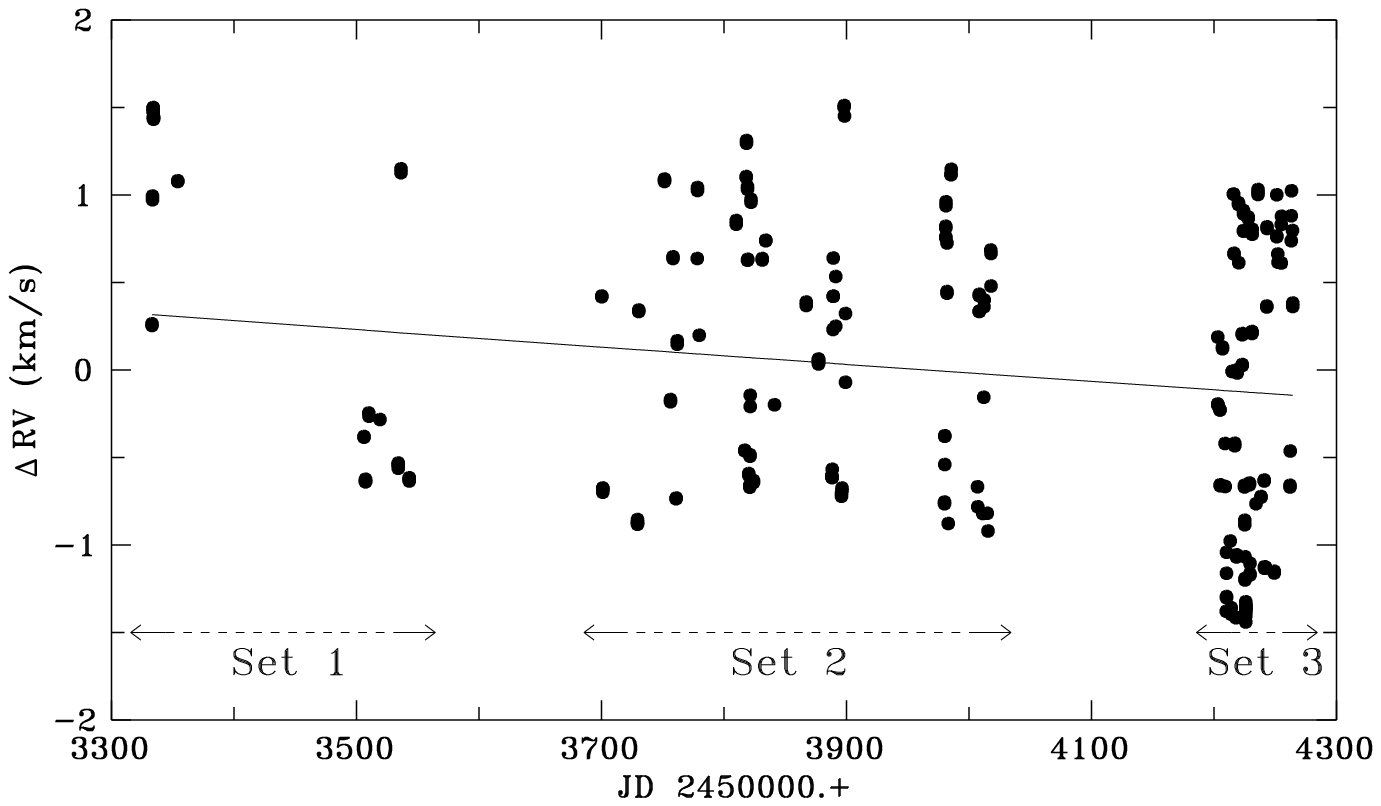}
   \caption{RV measurements of Polaris during 2004--2007.
   Solid line shows the decline of orbital RVs within interval of observations.
   }
\label{RV1}
\end{figure}

\clearpage

\begin{figure}
\centering
\includegraphics[width=16cm]{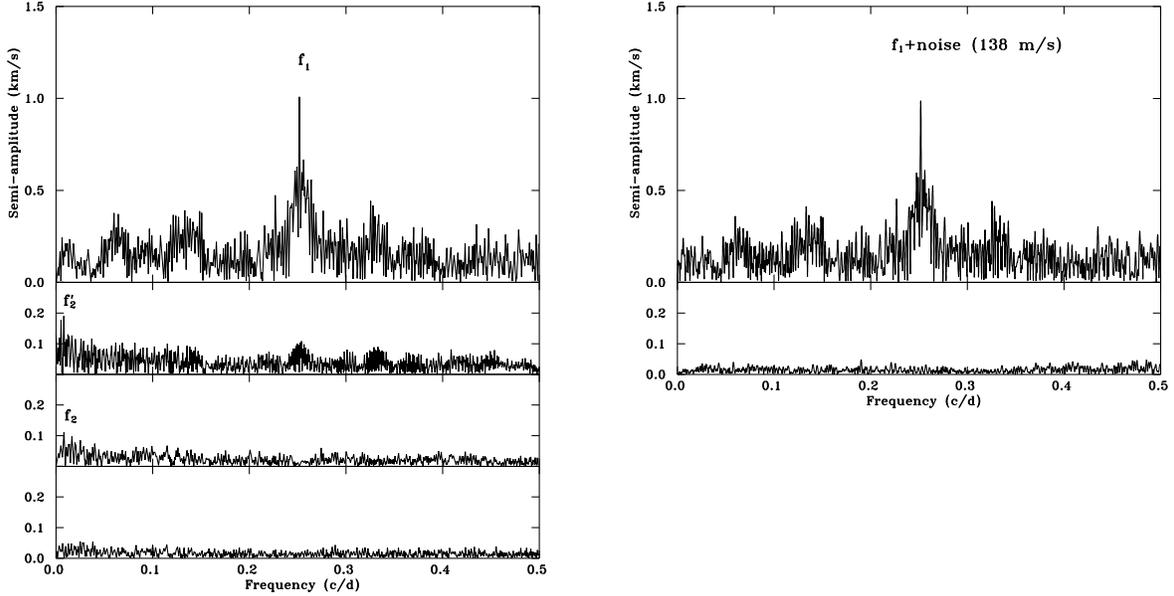}
   \caption{
   The amplitude spectra of DFT analysis for the entire RV measurements for Polaris.
   Left panel: (Top) DFT of original data. Largest peak is at $\it f_1$ = 0.251757 c d$^{-1}$.
   (Second) DFT of RV residuals after removing contribution from $\it f_1$.
   The second peak is at $f{_2}^{'}$ = 0.00799 c d$^{-1}$.
   (Third) DFT of merged residual RV  after removing of the best fit $\it f_1$
   contribution from individual  subsets (Set~1, Set~2, and Set~3).
   The secondary peak is at $f{_2}$ = 0.00840 c d$^{-1}$.
   (Bottom) DFT of residuals after removing of $f{_1}$ and $f{_2}$.
   Right panel: (Top) The amplitude spectrum of artificial
   sin-wave signal having the same frequency, amplitude and data point sampling as the
   dominant $f_1$ mode but co-added with normally distributed noise of
   the amplitude of 138 m/s. (Bottom) DFT of residual RV after
   removing contribution from $f_1$.
   }
\label{RV2}
\end{figure}

\clearpage

\begin{figure}
\centering
\includegraphics[width=8cm]{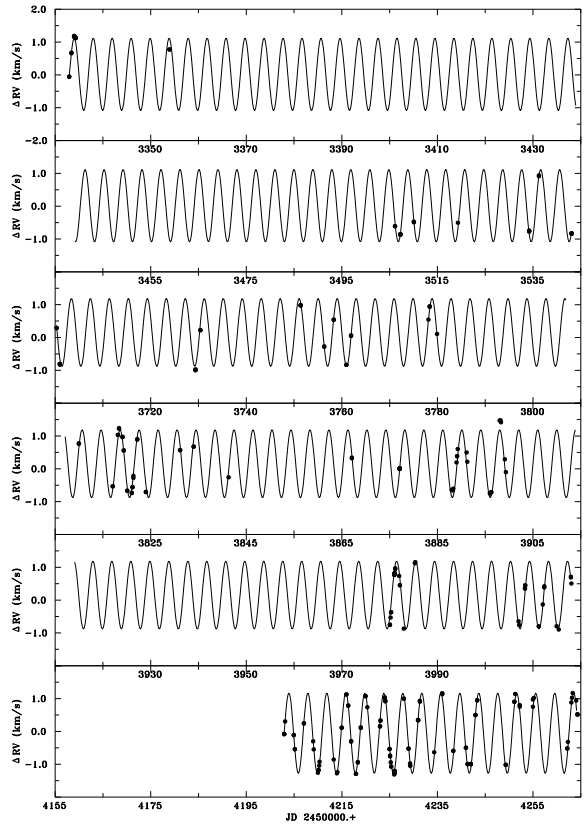}
\caption[]{Entire RVs of Polaris de-trended for orbital
variations. Solid line is a dominant period fit to every
subsets of the data.}
\label{Seg}
\end{figure}

\clearpage

\begin{figure}
\centering
\includegraphics[width=8cm]{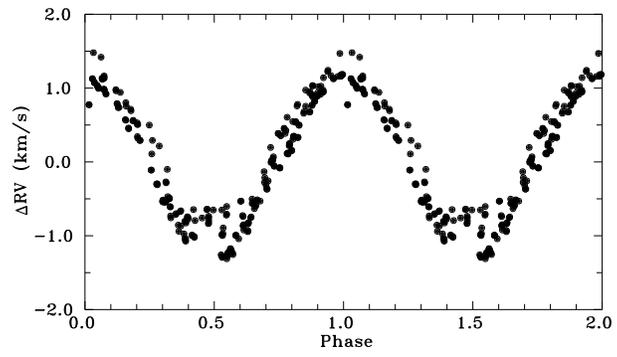}
   \caption{Phase curve of original data phased to the best fit
period $P_{1}$ = 3.97208 days.
   }
\label{RV3}
\end{figure}

\clearpage

\begin{figure}
\centering
\includegraphics[width=8cm]{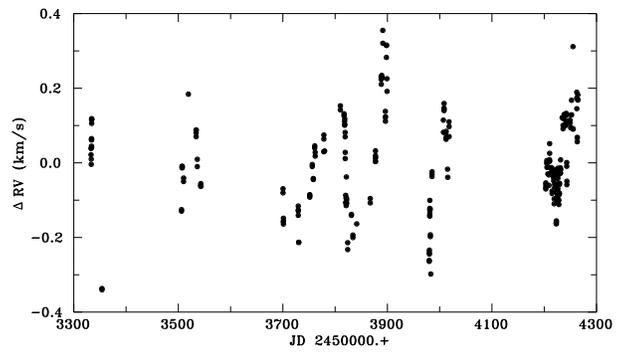}
\caption[]{
Merged residual RVs after removing the best fit
amplitude and period solutions to data from three subsets.
}
\label{Segm} \end{figure}

\clearpage

\begin{figure}
\centering
\includegraphics[width=8cm]{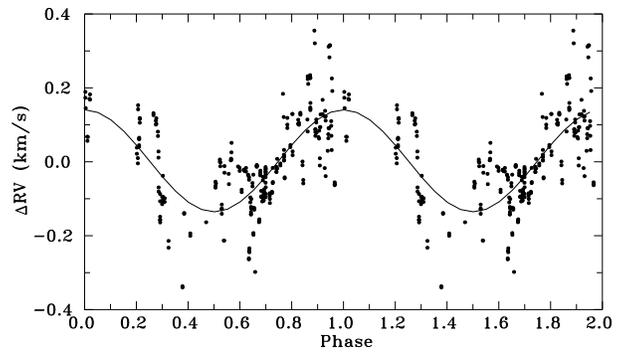}
\caption[]{Residual RV phased with a best fit period of $P_{2}$ = 119.1 day from three subsets.}
\label{Segm} \end{figure}

\clearpage

\begin{figure}
\centering
\includegraphics[width=10cm]{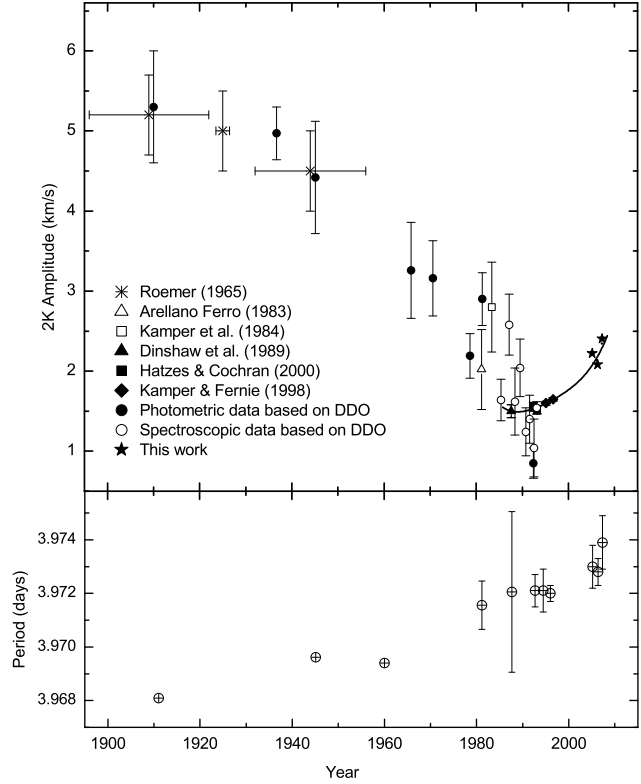}
   \caption{Variations of the pulsational RV amplitude and period of Polaris during the last century.
   (Top) Filled and open circles denote the result from DDO photometric and
   spectroscopic measurements, respectively.
   To convert photometric determination to RV determination, conversion factor of $\sim$ 50
   (Kamper $\&$ Fernie [1998] and Fernie et al. [1993]) was used.
   The solid line is a polynomial fit to the recent RV measurements.
   Error bars of Hatzes $\&$ Cochran (2000), Kamper $\&$ Fernie (1998), and this work are smaller than
   the symbols.
   (Bottom) The period variations;
   for the first three points, there are no reference in error bars.
   }
%   error bars for the recent three are smaller than symbols.

\label{RV6}
\end{figure}

\clearpage

\begin{table*}
\caption{The period and amplitude of Polaris.
$\sigma$ is r.m.s. residuals after main signal fitting
} \label{table:2}
\centering
\begin{tabular}{c c c c c c c}
\hline\hline
Data    & mean epoch      &  N & $\sigma$ & Semi-amplitude & Dominant Period& Period in residuals \\
    & (duration)      &  & (km s$^{-1}$)   &  (km s$^{-1}$) &  (days)   & (days) \\
\hline
Dinshaw & 1987.674 (0.660)  & 174   & 0.661 & 0.742 $\pm$ 0.068  & 3.97206 $\pm$ 0.00307 & 45   \\
Hatzes  & 1992.693 (1.698)  & 40    & 0.140 & 0.755 $\pm$ 0.032  & 3.97212 $\pm$ 0.00056 & 17, 40 \\
Kamper1 & 1994.492 (0.414)  & 71    & 0.098 & 0.786 $\pm$ 0.017  & 3.97208 $\pm$ 0.00081 &  34 \\
Kamper2 & 1995.973 (1.249)  & 129   & 1.107 & 0.825 $\pm$ 0.013  & 3.97200 $\pm$ 0.00029 &     \\
set 1   & 2005.183 (0.575)  & 34    & 0.111 & 1.105 $\pm$ 0.024  &
3.97300 $\pm$ 0.00080&$\backslash$    \\
set 2   & 2006.360 (0.870)  & 117   & 0.149 & 1.040 $\pm$ 0.021  &
3.97284 $\pm$ 0.00047&~~~~~~~~~$\rangle$ 119 \\
set 3   & 2007.349 (0.168)  & 113   & 0.063 & 1.203 $\pm$ 0.009  &
3.97394 $\pm$ 0.00098 &$/$   \\
\hline
%Entire data& 2453885.209          & 264    &   & 1.095 $\pm$ 0.018 & 3.9722006$\pm$ 0.00017 \\
%\hline
\end{tabular}
\end{table*}

\end{document}